\documentclass[11pt]{article}

\usepackage[T1]{fontenc}
\usepackage[utf8]{inputenc}
\usepackage[letterpaper,margin=1in]{geometry}
\usepackage{microtype}
\usepackage{amsmath,amssymb,amsthm}
\usepackage{graphicx}
\usepackage{booktabs}
\usepackage{multirow}
\usepackage{array}
\usepackage{subcaption}
\usepackage{makecell}
\usepackage[authoryear,round]{natbib}
\usepackage[section]{placeins}
\usepackage{xcolor}
\usepackage{xspace}
\usepackage{authblk}
\usepackage{hyperref}

\definecolor{citecolor}{rgb}{0.0,0.35,0.65}
\hypersetup{
  colorlinks=true,
  citecolor=citecolor,
  linkcolor=citecolor,
  urlcolor=citecolor,
  pdftitle={Scale-CDA: A Scalable Prototype to Democratize AI-Enabled, Cooperative Driving Automation for Production Vehicles},
  pdfauthor={Hao Zhou, Shengming Yuan, Yuhang Wang, Alina Hagen, Haibin Wen}
}
\graphicspath{{images/}}
\setcitestyle{authoryear,round}
\newcommand{\tech}{Scale-CDA\xspace}

\title{\tech: A Scalable Aftermarket Platform to Democratize Cooperative Driving Automation in Production Cars}
\author[1]{Hao Zhou\thanks{Corresponding author: \href{mailto:haozhou1@usf.edu}{haozhou1@usf.edu}}}
\author[1]{Shengming Yuan}
\author[1]{Yuhang Wang}
\author[1]{Alina Hagen}
\author[2]{Haibin Wen}
\affil[1]{University of South Florida, Tampa, FL, USA}
\affil[2]{Sunnypilot LLC, Virginia, USA}
\date{}

\begin{document}
\maketitle

\begin{abstract}

Scaling cooperative driving automation (CDA) from heavily-modified lab vehicles to general passenger cars requires an affordable, easy-to-install onboard unit (OBU) that can interface with existing OEM electronic systems in production vehicles. The central obstacle is OEM heterogeneity: Controller Area Network (CAN) signals and advanced driver-assistance system (ADAS) commands differ across car makes and models. Rather than re-engineering these interfaces for each vehicle, Scale-CDA builds upon OpenDBC, a community-maintained Python API to read-and-write hundreds of car models, and openpilot, an proven open-source Level-2 automation stack. Together, these ecosystems already provide access to L-2 automation across 300+ production car models. What remains missing for scalable CDA is a minimum-but-functional retrofit hardware platform, an additional connectivity layer, and Generative AI (GenAI) integration to enhance both connectivity and automation functions. This paper fills these gaps with Scale-CDA, an open-hardware and open-software CDA prototype that connects through a vehicle's CAN/ADAS interface. The prototype combines commodity edge computing, camera sensing, cellular and Wi-Fi communication, and a CAN adapter; a representative configuration costs less than \$1,000. It exchanges vehicle telemetry and cooperative messages using MQTT over Wi-Fi~6 or LTE, without requiring expensive DSRC or C-V2X sidelink radios. In a moving-vehicle experiment, Wi-Fi~6 achieved a mean message round-trip time of 5.25~ms and mean negotiated physical-layer rates of 98.51~Mb/s for transmission and 109.17~Mb/s for reception, although approximately 2\% of observations exceeded 50~ms. These results establish feasibility for non-safety-critical CDA applications. Scale-CDA also introduces a dual-purpose GenAI interface that combines camera observations, CAN streams, and received connectivity messages through the Model Context Protocol. The model produces semantic message intents and structured MetaActions; deterministic adapters validate these outputs, encode standards-compliant cooperative messages, and map admissible actions to existing Level-2 automation functions without granting the model direct actuator control. An on-road construction-zone demonstration exercises driver advisory, cooperative-message generation, and speed-related MetaAction pathways. Scale-CDA thus provides a reproducible path for adding connectivity and GenAI research capabilities to supported production vehicles.


\end{abstract}


\section{Introduction}




CDA improves an individual vehicle's decisions by incorporating information from other vehicles, roadway infrastructure, and cloud services. At a functional level, a CDA system must do three things: observe the vehicle and its surroundings, exchange relevant information with other participants, and translate the combined information into safe driving advice or actions. Existing research platforms provide these capabilities by adding specialized sensors, computing platforms, communication radios, and vehicle-specific control software to heavily modified experimental vehicles \citep{Mehr_2023,info16040317}. Although this approach is effective for demonstrating advanced functions, its cost and engineering burden make it difficult to replicate and impractical to extend across the production vehicles already on the road.

Scaling CDA to this existing fleet requires a different starting point: reuse the sensing, electronic control, and Level-2 automation capabilities that a vehicle already possesses, and add only the missing cooperative functions. This principle leads to three practical requirements. A retrofit CDA platform should be affordable, packaged as an aftermarket OBU with a repeatable installation procedure that trained automotive technicians can perform, and interoperable across vehicle manufacturers and models. Commodity cameras, edge computers, and wireless modems make such an affordable OBU feasible. Cross-vehicle interoperability is harder because CAN signals, ADAS commands, safety constraints, and physical connection methods differ among OEMs and even among models from the same OEM. Without a common vehicle interface and vehicle-specific installation procedures, every deployment would require a new reverse-engineering and integration effort.

\begin{figure}[!htbp]
    \centering
    \includegraphics[width=0.5\linewidth]{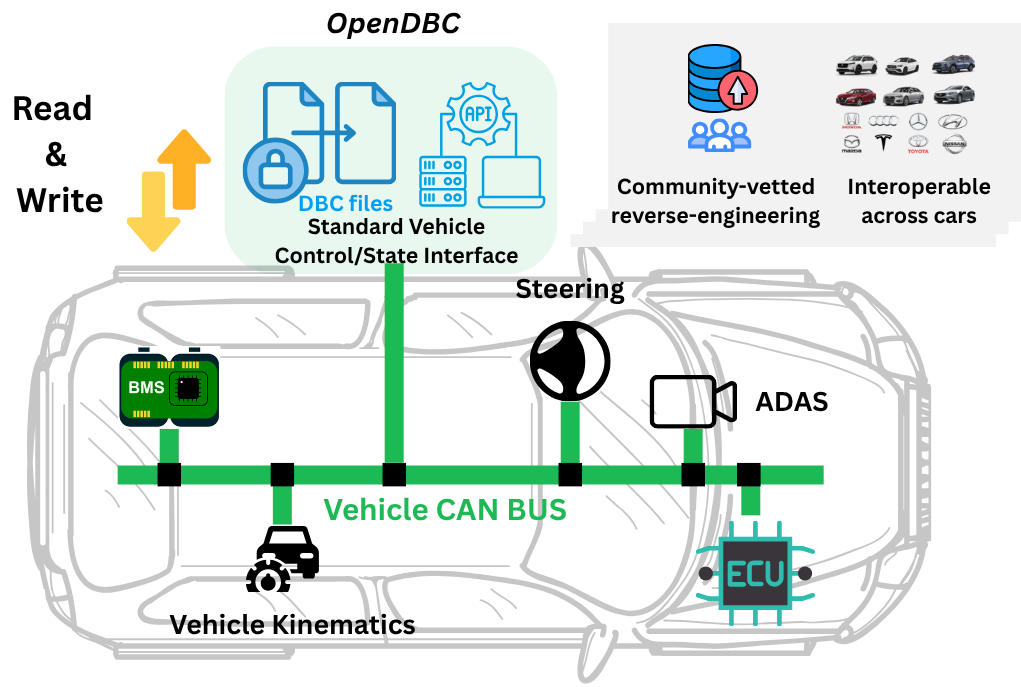}
    \caption{Overview of the OpenDBC vehicle-interface workflow}
    \label{fig:opendbc_function}
\end{figure}

The open-source OpenDBC and openpilot ecosystems provide this otherwise costly foundation. OpenDBC \citep{opendbc-git} is a community-maintained collection of vehicle-specific CAN definitions and Python interfaces that translate raw CAN traffic into meaningful vehicle states and supported commands. Openpilot \citep{openpilot-git} builds on these interfaces as an established Level-2 automation stack, providing perception, planning, driver monitoring, and bounded steering and longitudinal-control functions where the vehicle supports them. Together, the ecosystems cover more than 300 supported production-vehicle configurations. As summarized in Figure~\ref{fig:opendbc_function}, a CAN adapter connected through an accessible OBD-II or ADAS interface can use this software abstraction to read supported vehicle states and invoke permitted automation functions. A CDA platform can therefore inherit a cross-OEM vehicle interface instead of rebuilding one for every model.

Open vehicle interfaces and Level-2 automation alone, however, do not constitute CDA. They address observation and action within one vehicle but do not provide the cooperative exchange that distinguishes CDA. Three components are still needed: (i) retrofit hardware that combines computing, camera sensing, wireless communication, and vehicle access; (ii) a connectivity layer that exchanges vehicle telemetry and cooperative messages; and (iii) a controlled interface through which emerging decision-support methods, including GenAI, can consume multimodal context and interact with existing automation functions. Previous studies have used open vehicle interfaces for driving-control and data-collection research \citep{zhou2022incorporating,zhou2022congestion,wang2025openlkaopendatasetlane,wang2026adas}, but an integrated and reproducible path from this foundation to connectivity- and GenAI-enhanced CDA remains limited.

Connectivity presents the first integration challenge. Direct V2X technologies such as DSRC and C-V2X sidelink are designed for low-latency communication, but they require dedicated radios and supporting infrastructure \citep{s21030843}. Cellular and recent Wi-Fi technologies are more widely available and less expensive \citep{fi14100293}. They cannot be assumed to meet the reliability requirements of safety-critical V2X, but they may be adequate for less time-critical functions such as work-zone warnings, advisory speeds, traffic-management messages, and experimental data exchange. A scalable minimum platform should therefore determine what cooperative functions these commodity networks can support and characterize their limitations under motion rather than presume safety-critical performance.

GenAI presents the second integration challenge and an opportunity on both sides of CDA. Multimodal models can combine road images with structured vehicle and connectivity data to interpret situations that are difficult to describe using raw telemetry alone \citep{wen2023roadgpt4visionearlyexplorations,tang2025autoagentfullyautomatedzerocodeframework}. For communication, this reasoning can determine what contextual information is useful to share. For automation, it can select an appropriate high-level response. Yet neither output should be trusted as a packet or an actuator command: generative output does not guarantee protocol compliance, physical validity, or safe control. A practical interface must therefore separate reasoning from execution. GenAI may propose a semantic message or a bounded high-level action, while deterministic software validates and encodes the message, checks whether the action is admissible, and delegates planning and low-level control to the established automation stack.

This paper addresses these gaps through \tech, an exploratory open-hardware and open-software retrofit platform that adds a lightweight cooperative and GenAI layer to an established, OEM-interoperable Level-2 automation stack. Rather than proposing a new low-level controller, wireless standard, or foundation model, \tech integrates three reusable elements: an affordable and extensible OBU, Internet Protocol connectivity over Wi-Fi or cellular networks, and a dual-purpose GenAI interface for cooperative-message generation and bounded automation functions. The resulting architecture lets researchers replace or upgrade individual components while retaining the shared OpenDBC/openpilot vehicle interface, avoiding the need to construct a vehicle-specific CDA platform from the ground up.

Motivated by these research gaps, this paper makes the following three contributions:

\begin{enumerate}
    \item \textbf{An interoperable and affordable production-vehicle platform:}
    We develop a modular retrofit OBU that combines commodity edge computing, camera sensing, cellular and Wi-Fi communication, and a CAN interface with the OpenDBC and openpilot ecosystems. A representative configuration costs less than \$1,000 and can accommodate higher-performance computing and communication modules. Through the existing open ecosystems, \tech inherits vehicle-state and ADAS interfaces for 300+ car models.

    \item \textbf{A lightweight cellular and Wi-Fi connectivity framework:}
    We implement MQTT-based message exchange over Wi-Fi~6 and LTE without requiring dedicated DSRC or C-V2X sidelink radios. The framework carries vehicle telemetry and cooperative messages, including supported message structures based on SAE J2735 \citep{j2735}. A moving-vehicle experiment characterizes Wi-Fi~6 round-trip latency and negotiated physical-layer rates, establishing its feasibility and limitations for the evaluated non-safety-critical CDA applications.

    \item \textbf{GenAI interface to enhance both connectivity and automation:}
    We introduce a Model Context Protocol (MCP) interface through which a vision-language model combines camera observations with vehicle CAN data and received connectivity messages. For connectivity, the model proposes context-aware message intents that a deterministic adapter validates and encodes into standards-compliant cooperative-message structures. For automation, the model proposes structured MetaActions that are validated and mapped to existing Level-2 automation primitives rather than directly controlling vehicle actuators. An on-road construction-zone demonstration illustrates driver-advisory generation, cooperative-message generation, and the integration of a speed-related MetaAction.
\end{enumerate}

The remainder of the paper reviews related work, describes the hardware and software architecture, presents the connectivity and GenAI prototypes and demonstrations, and discusses applications, limitations, and future research needs.

\section{Related Work}

\subsection{CDA Platforms: From Simulation to Physical Deployment}

Cooperative driving automation has progressed from application-specific demonstrations toward reusable research ecosystems. SAE J3216 formalizes CDA as cooperation that supports or enables the dynamic driving task of a vehicle with an engaged driving-automation feature \citep{sae_j3216_2025}. The FHWA CARMA Platform operationalizes this concept through an open, ROS-based architecture with cooperative maneuver planning, vehicle and infrastructure messaging, and interfaces to host-vehicle hardware \citep{carma_platform}. OpenCDA complements CARMA with a modular full-stack environment for developing and evaluating cooperative perception, planning, and control in CARLA--SUMO co-simulation \citep{9564825,xu2023opencdaopensourceecosystemcooperative}. OpenCDA-ROS subsequently introduced a path between this simulation ecosystem and ROS-enabled physical vehicles and infrastructure \citep{zheng2023opencdaros}, while CDA.AI extends the ecosystem toward learning-based cooperative perception, prediction, planning, and system-level evaluation \citep{xu2025cdaai}. These developments establish that contemporary CDA research includes much more than isolated V2V applications: it now spans co-simulation, digital twins, cooperative sensing, maneuver coordination, and real-world prototyping.

The deployment assumptions of these platforms nevertheless differ from those of a broadly installable retrofit. CARMA is designed to be vehicle- and technology-agnostic above its hardware interface, but a physical host vehicle still requires compatible device drivers and vehicle-specific integration. OpenCDA and its extensions primarily support algorithm development in simulation or transfer to ROS-equipped research vehicles. Experimental platforms such as X-CAR similarly provide valuable full-stack access by extensively instrumenting a particular research vehicle \citep{Mehr_2023}. These approaches are appropriate for high-automation and algorithm-development research, but they do not by themselves provide an inexpensive interface to the sensing, CAN states, and bounded automation functions already present across many makes and models of production vehicles.

At the other end of the spectrum, commercial AI dashcams provide an easily installed combination of cameras, positioning, edge processing, and cellular connectivity, but representative products expose alerts, video, and fleet-management services through closed product ecosystems rather than an open vehicle-control and CDA research interface \citep{nexar_2021,gomotive.com_2022,samsara,connectedwise}. \tech targets the space between these two categories. It retains the packaging and commodity hardware of an aftermarket OBU, while building on OpenDBC and Openpilot to inherit a shared vehicle-state and Level-2 automation interface across more than 300 supported production-vehicle configurations \citep{opendbc-git,openpilot-git}. Its contribution is therefore not another full autonomous-driving stack; it is a lightweight cooperative layer that reuses an existing cross-OEM Level-2 foundation instead of reconstructing perception, planning, control, and CAN integration for every experimental vehicle.

\subsection{Connectivity: From Direct V2X Radios to Lightweight Networks}

Most safety-critical V2X research is appropriately organized around direct DSRC or C-V2X sidelink communication, for which bounded latency, reliability, congestion control, and standardized message exchange are central design requirements \citep{s21030843,s18051527}. CARMA, for example, includes encoders and decoders for SAE J2735 messages and uses additional mobility messages to support cooperative maneuver negotiation \citep{carma_platform,j2735}. This line of work provides the stronger communication foundation required for time-critical cooperative control, but dedicated radios, antennas, and roadside infrastructure increase the cost and engineering burden of small research deployments.

Internet Protocol connectivity offers a complementary deployment path. Prior studies have combined Wi-Fi and cellular links for V2X data exchange and have characterized newer Wi-Fi generations in outdoor environments \citep{8519489,9965626}. Commodity Wi-Fi and cellular networks cannot be presumed to provide the guarantees of a safety-critical sidelink, but they can support experimental telemetry and less time-critical applications such as work-zone advisories, traffic-management messages, and infrastructure data services. The remaining systems gap is to integrate these inexpensive links with both a production vehicle interface and recognizable CDA message semantics. \tech addresses that narrower gap using MQTT over Wi-Fi~6 or LTE as the transport and deterministic adapters for supported SAE J2735 message structures. It explicitly limits the evaluated links to non-safety-critical functions; thus, the design is a practical minimum CDA connectivity layer rather than a replacement for DSRC or C-V2X PC5.

\subsection{GenAI in Autonomous and Cooperative Driving}

GenAI integration into autonomous driving has developed rapidly beyond early natural-language demonstrations. GPT-Driver formulates motion planning as a language-model prediction problem, while early GPT-4V experiments examine the ability of general-purpose multimodal models to interpret road scenes and reason about driving decisions \citep{mao2023gptdriverlearningdrivegpt,wen2023roadgpt4visionearlyexplorations}. More recent systems move toward structured and closed-loop driving. LMDrive combines multimodal sensor data with language instructions for closed-loop end-to-end driving in CARLA \citep{shao2024lmdrive}; DriveLM uses graph-structured visual question answering to connect perception, prediction, and planning \citep{sima2024drivelm}; and DriveVLM decomposes scene description, analysis, and hierarchical planning, including a hybrid implementation that was demonstrated on a production vehicle \citep{tian2024drivevlm}. LangProp further explores using language models to generate and iteratively improve executable driving policies \citep{ishida2024langpropcodeoptimizationframework}. Collectively, this literature demonstrates concrete progress in scene understanding, long-tail reasoning, planning, human interaction, and closed-loop control, while also exposing persistent concerns about spatial reasoning, latency, hallucination, and verification \citep{cui2025largelanguagemodelsautonomous,drones9040238}.

GenAI has also begun to enter cooperative driving directly. V2X-VLM fuses vehicle- and infrastructure-side camera information with textual scene representations for end-to-end trajectory planning \citep{you2026v2xvlm}. V2V-LLM uses information from multiple connected vehicles for grounding, notable-object identification, and planning-oriented question answering \citep{chiu2025v2vllm}. Together with CDA.AI, these studies show that the intersection of GenAI and CDA is already an active research direction, particularly for cooperative perception and planning. The open question is consequently no longer whether a language or vision-language model can contribute to driving, but how to integrate that capability into deployable CDA without making unconstrained generative output part of a communication or actuation interface.

Existing GenAI-driving studies primarily evaluate a model, dataset, reasoning formulation, or end-to-end policy. Even cooperative VLM studies generally consume preassembled multi-vehicle or infrastructure sensor features and evaluate perception or trajectory outputs; generating, validating, and exchanging standards-structured cooperative messages is not their central focus. Conversely, most standard-oriented CDA platforms use deterministic perception and planning pipelines rather than a model-agnostic interface that supplies a GenAI model with synchronized camera observations, native vehicle states, and received cooperative messages. Evidence is also limited for carrying the same integration across heterogeneous production vehicles through an existing supervised Level-2 stack.

\subsection{Research Gap and Positioning of Scale-CDA}

The literature therefore leaves a specific integration gap between two mature but largely separate foundations: open CDA platforms provide standards-aware connectivity and cooperative applications on simulation or specialized research vehicles, while GenAI-driving systems provide increasingly capable multimodal reasoning and planning but are usually evaluated as model-centric or end-to-end autonomy solutions. What remains limited is \emph{an affordable, reproducible path that (i) runs on heterogeneous production vehicles, (ii) reuses their supported OEM interfaces and existing Level-2 functions, and (iii) uses GenAI to enhance both cooperative communication and automation without allowing free-form model output to become a packet or actuator command.}

\tech is distinctive at this systems boundary. Its model-agnostic interface combines camera observations, cross-OEM CAN states, and received connectivity messages. For standards-compliant cooperative messaging, GenAI proposes context-dependent semantic content, while deterministic software selects a supported message type, checks required fields and physical ranges, and serializes the result into the applicable SAE J2735 structure. For Level-2-ready automation, GenAI selects a structured MetaAction from a bounded vocabulary; a command bridge validates the request and maps it to an existing Openpilot planning or driver-interface function, leaving trajectory generation, low-level control, driver monitoring, and vehicle-specific safety enforcement in the established Level-2 stack. This division of responsibility makes the integration practical and replaceable: the model may run locally or in the cloud and may evolve independently, while the message and action boundaries remain explicit and testable. The reported prototype and on-road construction-zone demonstration establish functional feasibility of this architecture for supervised, non-safety-critical research; they do not establish GenAI as a safety-certified planner or \tech's commodity networks as substitutes for safety-critical V2X.

\section{Scale-CDA System Architecture}
\label{sec:system_architecture}

\begin{figure}[!htbp]
    \centering
    \includegraphics[width=\textwidth]{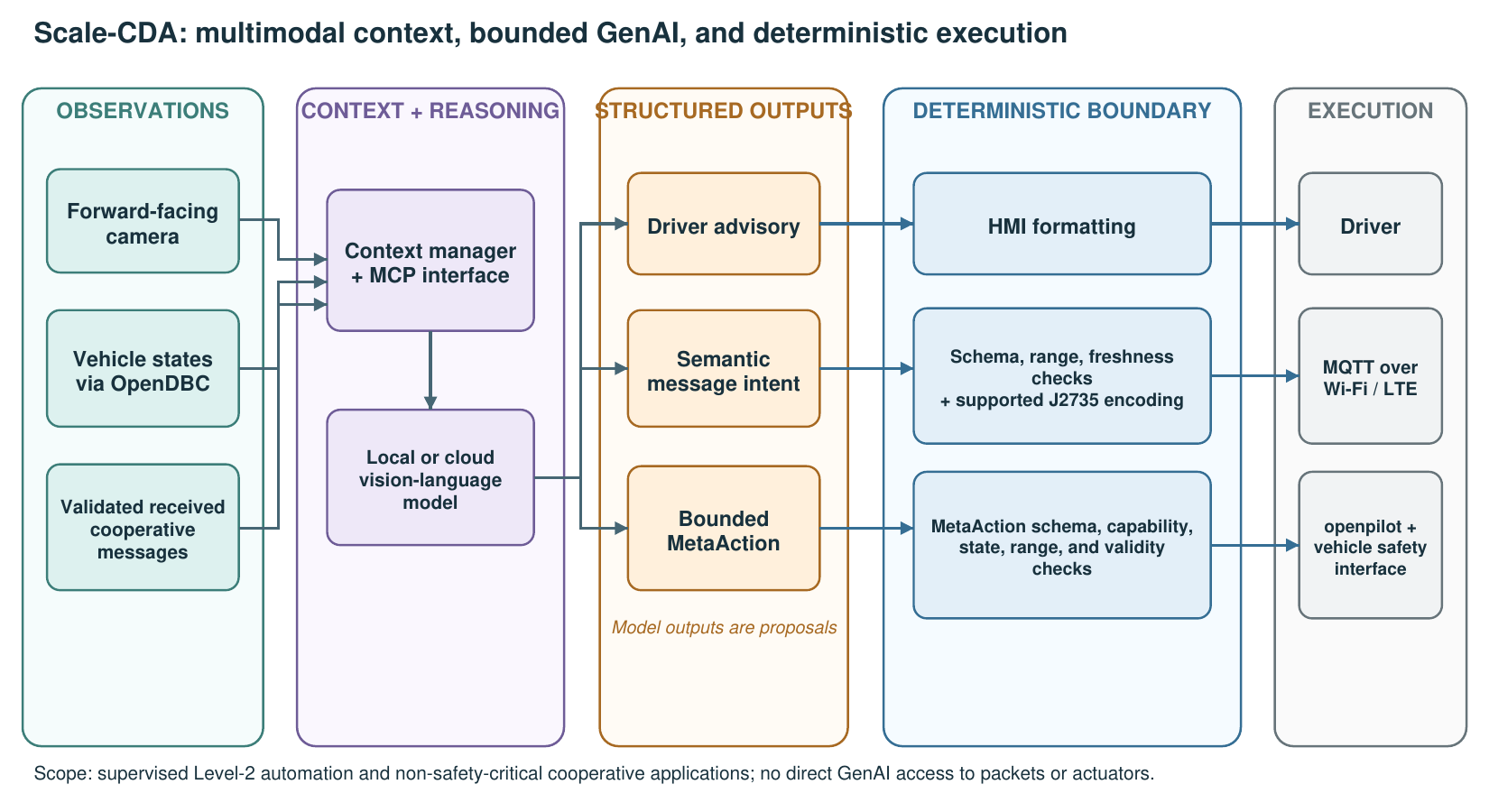}
    \caption{End-to-end Scale-CDA information flow. GenAI produces structured semantic outputs, while deterministic adapters retain responsibility for cooperative-message encoding and access to existing Level-2 automation functions.}
    \label{fig:system_architecture}
\end{figure}

Scale-CDA places a retrofit OBU between the vehicle-specific OpenDBC/openpilot interface and three higher-level functions: cooperative message exchange, multimodal GenAI reasoning, and supervised Level-2 automation. The architecture is organized around three boundaries. First, OpenDBC and openpilot isolate vehicle-specific CAN definitions, supported automation functions, and safety constraints from the cooperative applications. Second, MQTT over Wi-Fi or LTE carries experimental telemetry and cooperative messages without being treated as a safety-critical sidelink. Third, deterministic adapters separate GenAI reasoning from message serialization and vehicle execution. Figure~\ref{fig:system_architecture} summarizes these components and their information flow.

\subsection{Design Objectives and Operating Scope}

The architecture follows four design objectives derived from the deployment gap established in the preceding sections:
\begin{enumerate}
    \item \textbf{Affordable retrofit deployment:} the minimum OBU uses commodity computing, camera, vehicle-interface, and wireless components that can be installed without constructing a purpose-built research vehicle.
    \item \textbf{Cross-OEM reuse:} vehicle-specific signal decoding and permitted automation commands are inherited from OpenDBC and openpilot rather than reimplemented for each make and model.
    \item \textbf{Modular connectivity:} cooperative applications use an application-layer interface that is independent of whether the IP connection is provided by Wi-Fi or LTE.
    \item \textbf{Bounded GenAI integration:} a vision-language model may interpret multimodal context and propose semantic outputs, but deterministic software retains responsibility for message encoding, validity checks, and access to Level-2 automation functions.
\end{enumerate}

Scale-CDA is an experimental platform for supervised Level-2 and non-safety-critical CDA applications. The commodity networks evaluated in this work do not provide the bounded latency and reliability required of safety-critical V2X, and the GenAI layer is not part of the low-level control loop. The driver remains responsible for supervising the engaged automation feature. If a generated output is malformed, stale, unsupported, or outside an application-defined range, the corresponding adapter rejects it without transmitting a message or changing an automation setting.

\subsection{End-to-End Information Flow}

The OBU collects three forms of context: forward-facing camera observations, vehicle states decoded from the CAN bus, and the latest cooperative messages received through the connectivity module. A context manager attaches source and time information to these inputs and exposes the available data through an MCP interface. MCP standardizes model access to the structured context; it does not itself perform perception, planning, message encoding, or control. Freshness limits are enforced when the data are consumed so that an old vehicle state or connectivity message is not silently treated as a current observation.

The inference backend is model-agnostic and may be local or cloud-hosted. It produces three typed outputs rather than unrestricted commands: a driver advisory, a semantic message intent, and a MetaAction selected from a bounded vocabulary. Each output follows a separate downstream path. Advisories are presented through the human--machine interface. Message intents pass through schema and physical-range validation before a supported cooperative-message structure is serialized and published. MetaActions pass through capability, automation-state, parameter-range, and validity checks before an admissible request is mapped to an existing openpilot planning or driver-interface function. The openpilot stack remains responsible for trajectory generation, low-level control, driver monitoring, and vehicle-specific safety enforcement.

\subsection{Retrofit OBU and Vehicle Abstraction}

\subsubsection{OBU Hardware}

The representative OBU combines an x86 edge computer, a forward-facing camera, a CAN adapter and vehicle harness, and Wi-Fi and LTE interfaces. A positioning source can provide location for applications that require it, while an external display or the existing vehicle interface can present driver advisories. In the reported prototype, the edge computer hosts the Scale-CDA context, connectivity, message-adapter, and MetaAction-bridge services. GenAI inference can run on the same computer when the selected model and latency requirement permit, or it can be delegated to a cloud service through the same structured interface.

\begin{figure}[!htbp]
    \centering
    \begin{subfigure}{0.44\textwidth}
        \includegraphics[width=\linewidth]{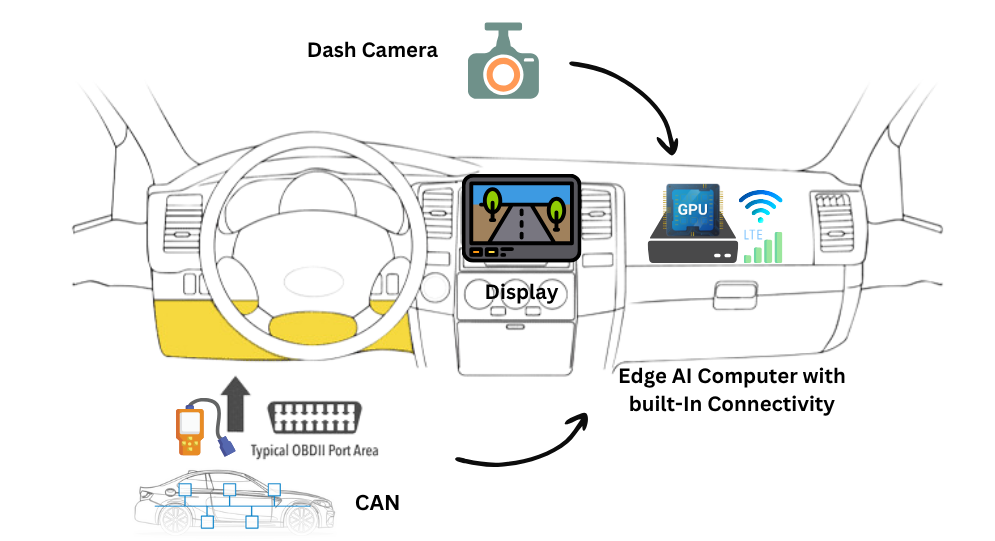}
        \caption{Functional placement in the vehicle}
    \end{subfigure}
    \hfill
    \begin{subfigure}{0.44\textwidth}
        \includegraphics[width=\linewidth]{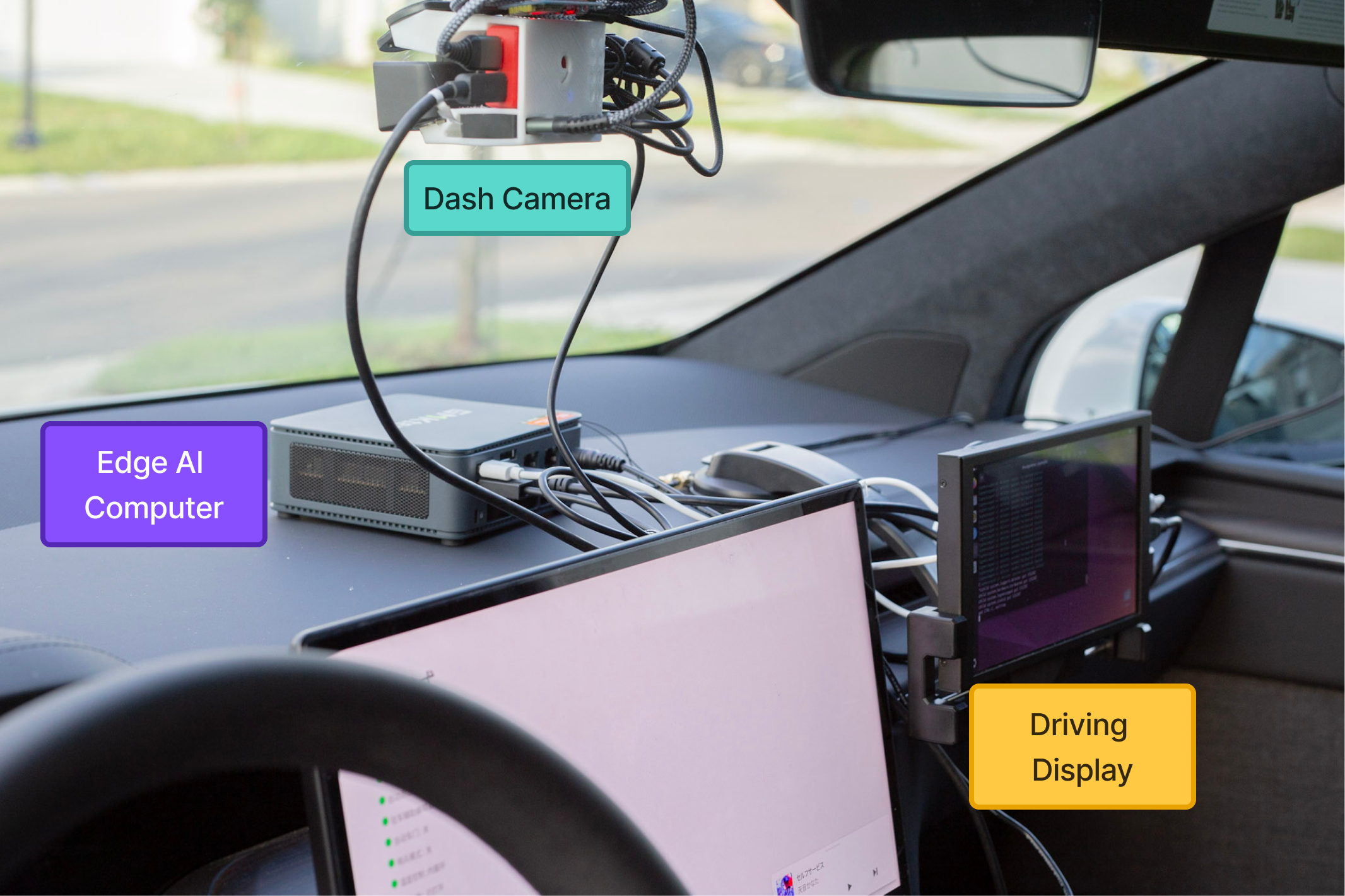}
        \caption{Prototype installed in a production vehicle}
    \end{subfigure}
    \caption{Scale-CDA retrofit OBU hardware.}
    \label{fig:hardware_overview}
\end{figure}

The camera in the reported GenAI pipeline supplies contextual road-scene images. It is not treated as an independently certified safety sensor, and GenAI interpretation of its images does not replace the perception, planning, and driver-monitoring responsibilities of the supervised Level-2 stack. Figure~\ref{fig:hardware_overview} shows the functional placement and physical assembly of the prototype OBU.

Table~\ref{tab:hardware_cost} reports the representative component prices used to substantiate the sub-\$1,000 core-OBU claim. The listed configuration totals approximately \$820 before taxes and excludes an optional external display and vehicle-specific installation materials. Higher-performance compute devices can be substituted without changing the software interfaces.

\begin{table}[!htbp]
\centering
\caption{Representative Scale-CDA core-OBU component cost.}
\label{tab:hardware_cost}
\begin{tabular}{p{0.29\linewidth}p{0.48\linewidth}r}
\toprule
\textbf{Component} & \textbf{Role in the prototype} & \textbf{Cost (USD)} \\
\midrule
GMKtec NucBox K6 & Context, connectivity, automation-integration, and optional local-inference services & \$480 \\
Standard webcam & Forward-facing observations for the GenAI context pipeline & \$40 \\
CAN adapter and vehicle interface & Access to supported vehicle states and automation functions & \$100 \\
LTE and Wi-Fi interfaces & IP connectivity for cooperative applications & \$200 \\
\midrule
\textbf{Representative total} & & \textbf{\$820} \\
\bottomrule
\end{tabular}
\end{table}

\subsubsection{Cross-OEM Vehicle Interface}

The vehicle interface consists of a CAN adapter, an accessible OBD-II or ADAS connection, and the applicable OpenDBC/openpilot software definition. OpenDBC translates supported raw CAN signals into common vehicle-state fields such as ego speed, steering state, pedal status, and cruise settings. Openpilot adds perception, planning, driver monitoring, and bounded steering or longitudinal-control functions where they are supported by the vehicle configuration. Scale-CDA consumes these shared interfaces rather than embedding OEM-specific CAN identifiers in its cooperative or GenAI modules.

Physical installation remains a nontrivial part of this interface. For vehicle configurations that use an ADAS-module connection, a technician must access the module enclosure, disconnect the OEM harness, insert the vehicle-specific Scale-CDA harness, reassemble the enclosure, and perform post-installation diagnostic and functional checks. The electrical interface is designed to avoid cutting or permanently altering the factory wiring, but the access procedure, connector location, and verification steps vary by vehicle configuration. Scalable deployment therefore requires documented vehicle-specific procedures and technician training in addition to software compatibility.

Compatibility does not imply identical functionality on every supported vehicle. The available signals, connection method, and steering or longitudinal-control capabilities vary by model, trim, and OEM safety design. The command bridge therefore uses the active vehicle configuration to expose only supported states and MetaActions and leaves vehicle-specific command checks within the established interface. Neither an incoming cooperative message nor a GenAI response is forwarded directly to the CAN bus.

\subsection{Lightweight Cooperative Connectivity}

The connectivity architecture separates message semantics from network transport. At the application layer, Scale-CDA handles vehicle telemetry, received cooperative information, and semantic message intents. A deterministic adapter selects an implemented message type, adds system-derived fields such as the vehicle identifier, location, and timestamp, validates required values and units, and serializes the result into the supported SAE J2735-based structure. Unsupported or invalid intents are rejected. This design does not claim implementation of the full J2735 message set.

At the transport layer, an MQTT client publishes and subscribes through either a locally reachable broker over Wi-Fi or a remote broker over LTE. Broker-mediated IP communication supports V2I, V2N, and experimental V2V information exchange, but it is distinct from direct DSRC or C-V2X sidelink communication. Topic organization, MQTT quality-of-service level, and broker placement can be configured for the application; these settings do not convert the underlying commodity network into a safety-guaranteed channel. Incoming messages are decoded and checked for type, format, and freshness before they are exposed to an application or the MCP context interface.

Authentication, transport encryption, broker authorization, and credential management are necessary for deployment beyond a controlled research setting. These security mechanisms can be added at the MQTT and network layers, but they are not evaluated by the communication experiment in Section~\ref{sec:connectivity}. The present evaluation therefore concerns functional exchange and observed network performance, not security certification or safety-critical communication guarantees.

\subsection{Controlled GenAI Integration}

The GenAI layer is designed as a replaceable reasoning service rather than a controller. Through MCP, the model can access camera observations, decoded vehicle states, and validated cooperative messages without receiving raw CAN write access. A local backend can reduce data exposure and dependence on wide-area connectivity, whereas a cloud backend can provide greater model capacity at the cost of network dependence and transmission of selected context outside the vehicle. Both deployment modes use the same input and output contracts.

The model returns a structured object with up to three fields: \texttt{driver\_advisory}, \texttt{message\_intent}, and \texttt{meta\_action}. The first is an informational output for the driver. The second describes what context may be useful to communicate but leaves message-type selection, system-derived fields, validation, and serialization to the deterministic message adapter. The third names one supported high-level automation function and its bounded parameters; it does not contain source code, a trajectory, or raw steering, braking, or acceleration commands.

The MetaAction bridge validates the action schema, parameter units and ranges, validity interval, current automation state, and vehicle capability. An accepted action is mapped to the applicable openpilot planning, state-machine, or driver-interface function. A rejected action has no effect on the automation stack. Thus, the model may propose \emph{what} the system should communicate or which supported high-level response to consider, while conventional software determines whether and how that proposal is executed.

This separation also preserves modularity. The camera, wireless interface, model backend, supported message set, and MetaAction library can evolve independently, while the OpenDBC/openpilot vehicle boundary and deterministic validation paths remain stable. 

\section{Prototype Implementation and Demonstrations}
\label{sec:prototype}

This section describes the implementation used to exercise the architecture in Section~\ref{sec:system_architecture}. The evaluation has three purposes: select a practical edge-computing platform, characterize the implemented Wi-Fi message path under vehicle motion, and demonstrate that one multimodal observation can drive the advisory, cooperative-message, and bounded MetaAction pathways. The experiments establish functional feasibility of the prototype; they are not a safety assessment or a comparative evaluation of CDA communication and decision-making technologies.

\subsection{Edge-Computing Platform Selection}
\label{hardware_val}

We screened three commodity edge-computing platforms: an NVIDIA Jetson AGX Orin, an NVIDIA Jetson Orin Nano \citep{NVIDIAJetsonOrin}, and a GMKtec NucBox K6 equipped with an AMD Ryzen~7 7840HS CPU and integrated Radeon 780M GPU \citep{gmktex_2025,amd_7840}. The selection considered acquisition cost, development environment, support for the vehicle and connectivity software, and local VLM decoding throughput. LLaVA-7B \citep{liu2023visualinstructiontuning} and Gemma~3-4B \citep{gemma3report} were used as representative local models.

Table~\ref{tab:hardware_benchmark} reports the measured generated-token rates in the prototype screening. These values characterize decoding throughput for the tested model--device combinations; they do not include image preparation, prompt prefill, network transfer, output parsing, or downstream validation and therefore should not be interpreted as end-to-end advisory latency.

\begin{table}[!htbp]
\centering
\caption{Exploratory local VLM decoding throughput used for platform selection.}
\label{tab:hardware_benchmark}
\begin{tabular}{@{}lrrr@{}}
\toprule
\textbf{Device} & \textbf{Approx. cost} & \textbf{LLaVA-7B} & \textbf{Gemma 3-4B} \\
& \textbf{(USD)} & \textbf{(tokens/s)} & \textbf{(tokens/s)} \\
\midrule
NVIDIA Jetson AGX Orin & \$1,999 & 28.62 & 28.62 \\
NVIDIA Jetson Orin Nano & \$250 & 7.17 & 6.27 \\
GMKtec NucBox K6 & \$480 & 12.04 & 13.65 \\
\bottomrule
\end{tabular}
\end{table}

The NucBox K6 was selected for the reported OBU because it provided a standard x86 development environment and intermediate local-inference throughput at substantially lower cost than the AGX Orin. In the prototype, it hosts the vehicle-context, MQTT, message-adapter, and MetaAction-bridge services. Local decoding on the screened devices did not make the complete multimodal advisory pipeline sufficiently responsive for the reported on-road use case, so the on-road demonstration used a cloud-hosted Qwen3-VL model \citep{qwen3vlreport}. The architecture nevertheless retains the local-inference option for smaller or optimized models and for applications with less stringent response-time requirements.

\begin{figure}[!htbp]
    \centering
    \includegraphics[width=0.98\linewidth]{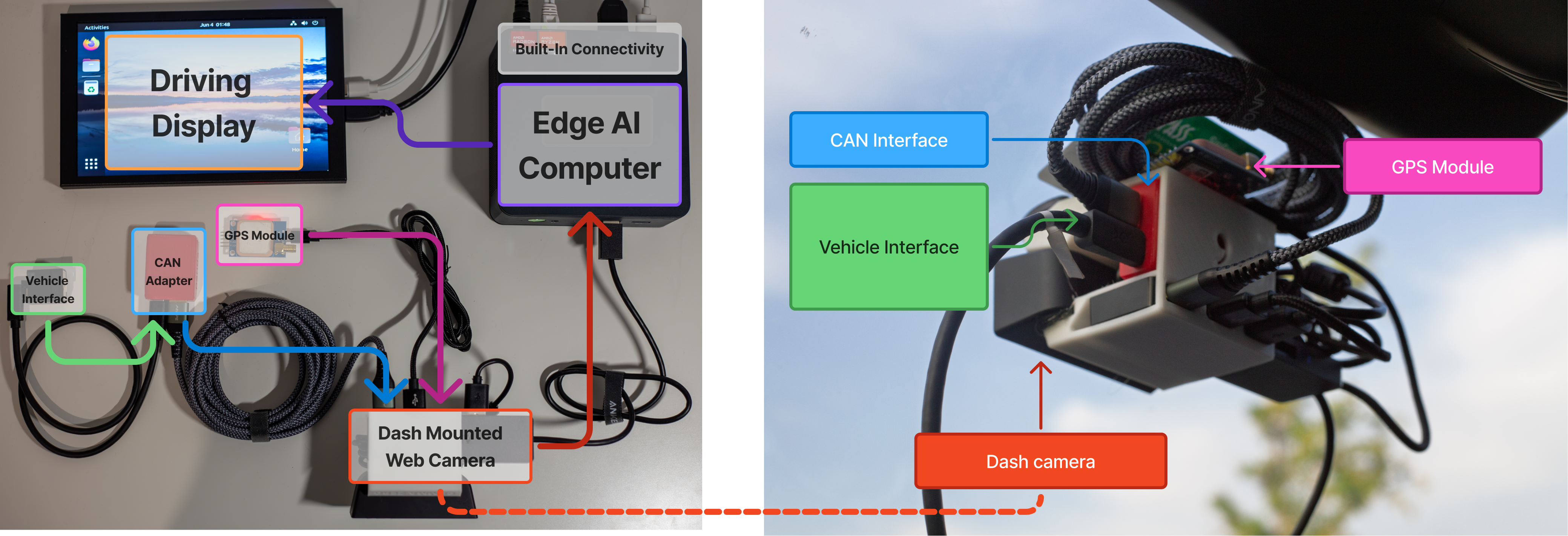}
    \caption{Physical hardware connections and sensor assembly of the Scale-CDA prototype.}
    \label{fig:sensor_suites}
\end{figure}

Figure~\ref{fig:sensor_suites} documents the physical prototype corresponding to the logical OBU architecture in Figure~\ref{fig:hardware_overview}. The bench layout shows the data and display connections among the vehicle interface, CAN adapter, positioning source, dash-mounted camera, edge computer, and driver display. The in-vehicle view shows the compact camera, CAN-interface, and positioning assembly used during deployment. This figure describes the implemented wiring and packaging rather than an additional software pathway.

\subsection{MQTT Connectivity Implementation}
\label{sec:connectivity}

The OBU runs an MQTT client that publishes vehicle telemetry and validated cooperative-message payloads and subscribes to messages from other clients. MQTT is a lightweight publish--subscribe protocol widely used in Internet-of-Things systems \citep{soni2017survey,app9050848}. The prototype used HiveMQ \citep{hivemq_2025} as the broker. For the local Wi-Fi configuration, the broker and echo service were reachable through the roadside network; for LTE operation, the same application interface could connect to a remote broker. The Wi-Fi experiment below evaluates the local path only. It does not provide a measured comparison with LTE.

\subsubsection{Moving-Vehicle Wi-Fi Experiment}
\label{wifi_testing}

The experiment was designed to characterize application-layer message RTT and the negotiated physical-layer link rate while the OBU was moving. Two Ubiquiti U7 Outdoor access points \citep{u7outdoorspec} were mounted approximately 2~m above ground and positioned 50~m apart along a roughly 70~m parking-lot lane. Each access point used its 5-GHz radio, a 20-MHz channel, a configured transmit power of 26~dBm, and an omnidirectional antenna with 4~dBi gain. The vehicle used a MediaTek MT7922 Wi-Fi~6/6E client \citep{mediatek_mt7922} connected to a 9-dBi omnidirectional antenna mounted on the passenger side.

\begin{figure}[!htbp]
    \centering
    \begin{subfigure}[t]{0.62\textwidth}
        \includegraphics[width=\linewidth]{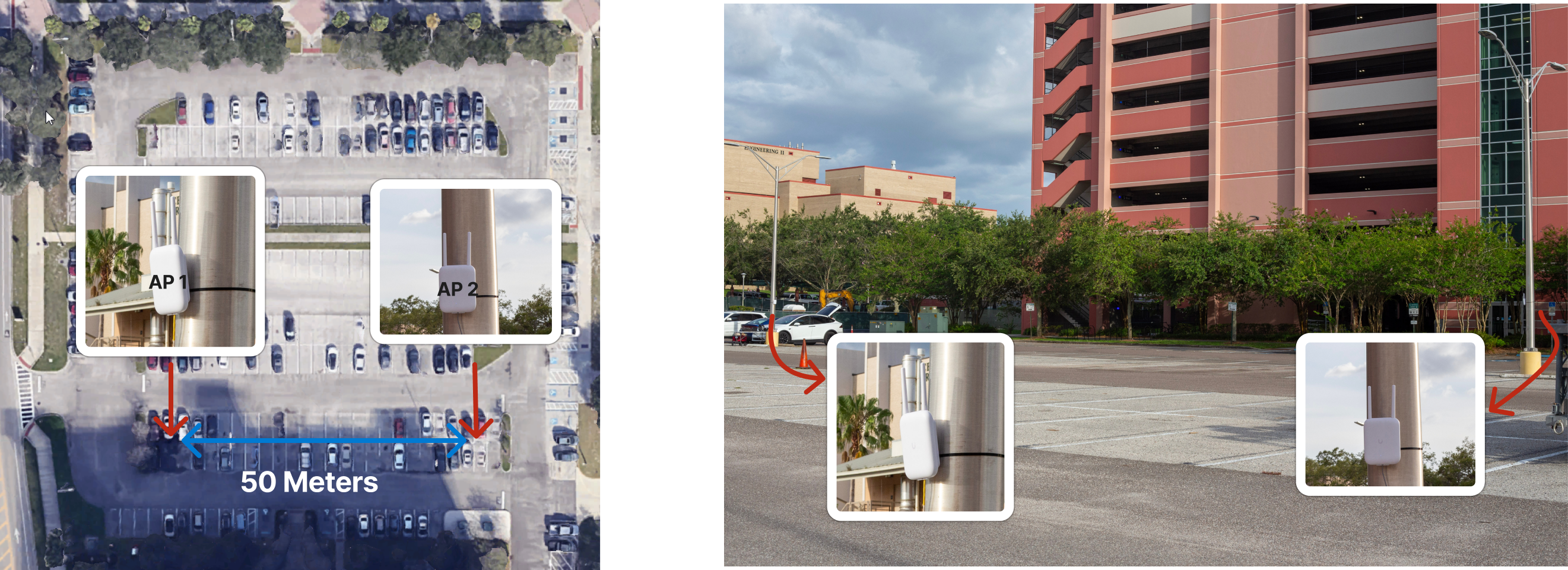}
        \caption{Access-point placement and field installation}
    \end{subfigure}
    \hfill
    \begin{subfigure}[t]{0.32\textwidth}
        \includegraphics[width=\linewidth]{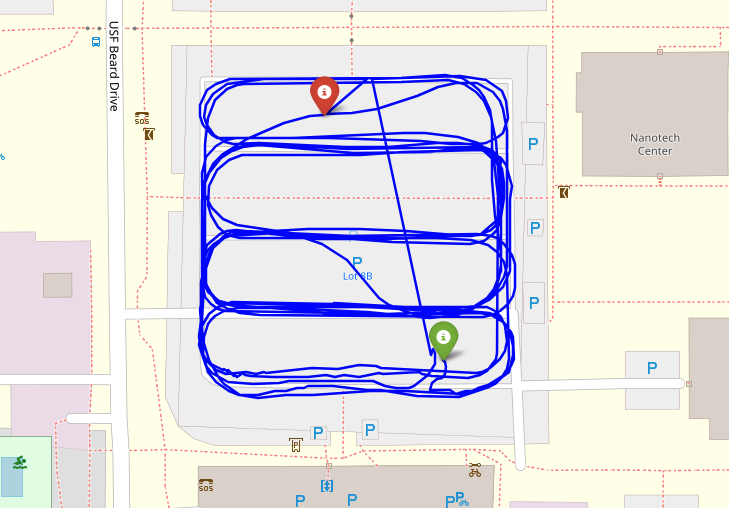}
        \caption{Recorded vehicle path}
    \end{subfigure}
    \caption{Moving-vehicle Wi-Fi experiment. The two access points were separated by 50~m; the vehicle completed repeated loops through the surrounding parking lot.}
    \label{fig:wifi_experiment}
\end{figure}

The vehicle completed repeated loops at a mean speed of 14.40~km/h. The test client transmitted payloads averaging 190~bytes to the local server, which returned each payload to the OBU for RTT measurement. The logged dataset contains 3,600 message exchanges. The OBU remained associated with the Wi-Fi network throughout the recorded run; this observation does not by itself establish packet-delivery reliability.

\begin{figure}[!htbp]
    \centering
    \includegraphics[width=0.98\linewidth]{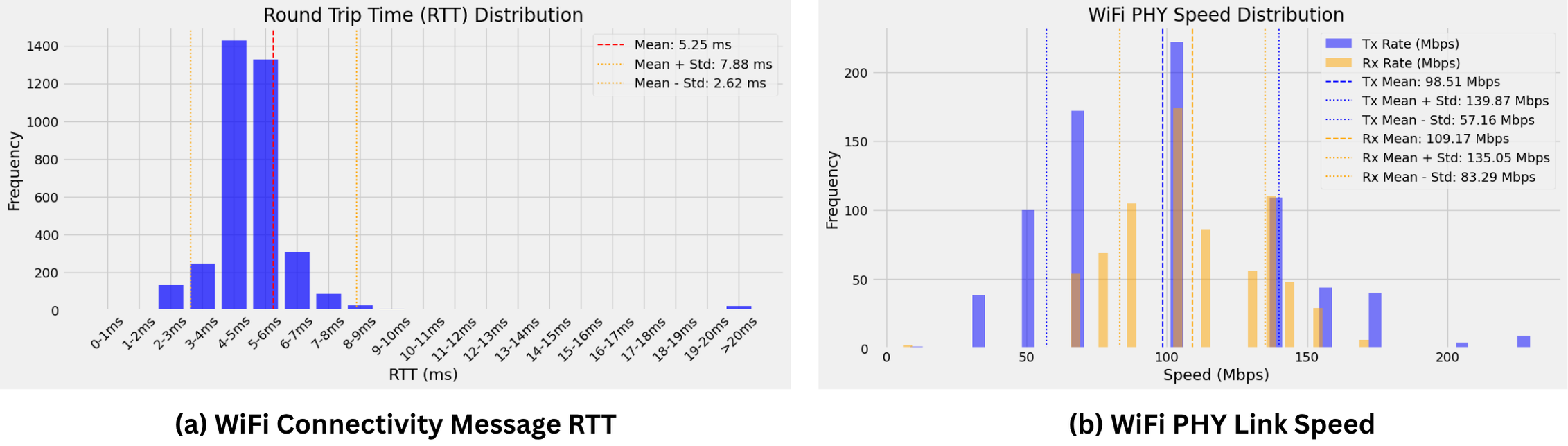}
    \caption{Moving-vehicle Wi-Fi results: (a) application-layer message RTT and (b) negotiated transmit and receive physical-layer rates. The final RTT bin aggregates observations above 20~ms; the fraction above 50~ms was calculated from the recorded samples.}
    \label{fig:wifi_rtt}
\end{figure}

Figure~\ref{fig:wifi_rtt} shows the RTT and negotiated link-rate distributions, and Table~\ref{tab:wifi_results} summarizes the reported measurements. Mean RTT was 5.25~ms, although approximately 2\% of observations exceeded 50~ms. Mean negotiated rates were 98.51~Mb/s for transmission and 109.17~Mb/s for reception. These physical-layer rates describe the link mode selected by the devices and are not equivalent to measured application throughput.

\begin{table}[!htbp]
\centering
\caption{Summary of the moving-vehicle Wi-Fi measurements.}
\label{tab:wifi_results}
\begin{tabular}{@{}p{0.43\linewidth}p{0.20\linewidth}p{0.27\linewidth}@{}}
\toprule
\textbf{Metric} & \textbf{Result} & \textbf{Interpretation} \\
\midrule
Mean message RTT & 5.25~ms & Application-layer echo RTT \\
RTT observations above 50~ms & approximately 2\% & Observed long-latency tail \\
Mean negotiated transmit rate & 98.51~Mb/s & Physical-layer rate, not throughput \\
Mean negotiated receive rate & 109.17~Mb/s & Physical-layer rate, not throughput \\
\bottomrule
\end{tabular}
\end{table}

The low mean RTT supports feasibility for the evaluated telemetry and advisory applications, but the long-latency tail prevents treating the link as deterministic. The experiment was conducted at one site, at low vehicle speed, with two access points and a small payload. It did not independently measure loss, handoff interruption, interference sensitivity, coverage probability, MQTT quality-of-service effects, or end-to-end application deadlines. The results therefore do not establish safety-critical suitability, broad-area coverage, or superiority over LTE, legacy Wi-Fi, DSRC, or C-V2X.

\subsubsection{Cooperative-Message Path}

The MQTT implementation carries two payload families. The first contains decoded vehicle telemetry for experimental monitoring and data exchange. The second contains cooperative information produced by deterministic application logic or from a validated GenAI message intent. A GenAI response is not transmitted directly. The message adapter selects a supported message mapping, adds system-derived fields such as vehicle identifier, location, and timestamp, checks required fields, types, units, ranges, and freshness, and then serializes the result for MQTT publication.

For mappings implemented by the prototype, the adapter targets the applicable SAE J2735-based structure \citep{j2735}. Unsupported, incomplete, stale, or out-of-range intents are rejected. This work does not implement the full J2735 message dictionary or present an independent conformance test; its claim is limited to the supported structures exercised by the prototype. The construction-zone demonstration below exercises the semantic-intent, validation, serialization, and publication pathway.

\subsection{GenAI and MetaAction Implementation}
\label{sec:genai_implementation}

The GenAI prototype implements the three-input context interface defined in Section~\ref{sec:system_architecture}. A forward-facing camera supplies road-scene images; OpenDBC supplies decoded vehicle states such as ego speed, pedal status, steering state, and cruise setting; and the MQTT subscriber supplies the latest validated cooperative messages. Vehicle and connectivity data are exposed through MCP \citep{anthropic_mcp}. The reported on-road demonstration used a cloud-hosted Qwen3-VL model, while the local-device screening in Section~\ref{hardware_val} evaluated the alternative onboard deployment.

An \textit{openEMMA}-inspired task-decomposition prompt \citep{xing2025openemmaopensourcemultimodalmodel} guides the model through scene description, identification of relevant road users, interpretation of the structured context, and response generation. The application requests a JSON object with three optional outputs. Table~\ref{tab:genai_outputs} identifies each output and its deterministic consumer.

\begin{table}[!htbp]
\centering
\caption{Structured GenAI outputs and downstream processing in the prototype.}
\label{tab:genai_outputs}
\begin{tabular}{@{}p{0.19\linewidth}p{0.29\linewidth}p{0.42\linewidth}@{}}
\toprule
\textbf{Output} & \textbf{Purpose} & \textbf{Downstream processing} \\
\midrule
\texttt{driver\_advisory} & Natural-language information for the driver & Text-format checks and presentation through the HMI \\
\texttt{message\_intent} & Semantic content proposed for cooperative exchange & Supported-event, field, unit, range, freshness, and message-mapping checks \\
\texttt{meta\_action} & Requested high-level Level-2 interaction & Action allowlist, vehicle capability, automation-state, parameter, and validity checks \\
\bottomrule
\end{tabular}
\end{table}

The three fields are consumed independently. Failure of one pathway does not require the other two to be accepted. In particular, free-form explanatory text is never parsed as a network packet or actuator command.

\begin{figure}[!htbp]
    \centering
    \includegraphics[width=\textwidth]{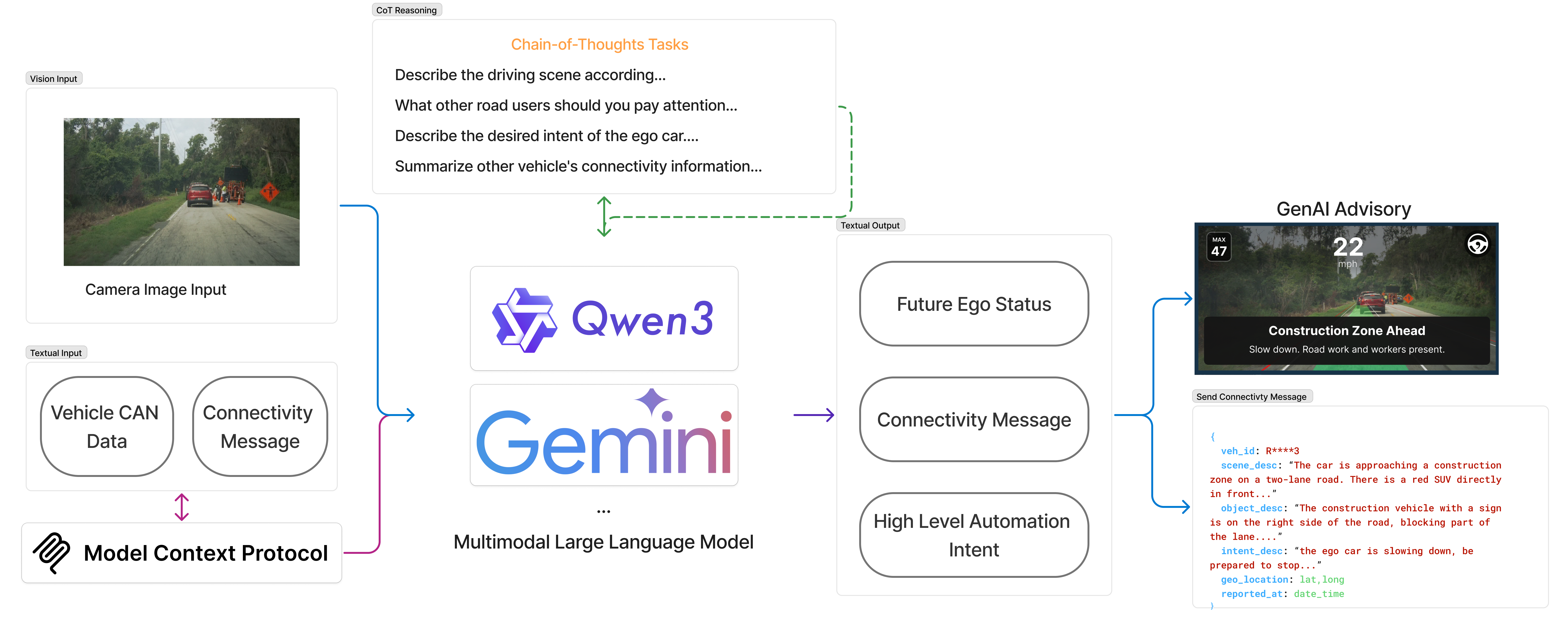}
    \caption{Implementation-level GenAI workflow for producing a driver advisory, semantic cooperative-message intent, and high-level automation intent from camera, vehicle, and connectivity context. The three conceptual outputs are represented in the current implementation by \texttt{driver\_advisory}, \texttt{message\_intent}, and \texttt{meta\_action}; model-generated message content is validated and encoded by deterministic software before publication.}
    \label{fig:genai_workflow}
\end{figure}

Figure~\ref{fig:genai_workflow} provides an implementation-level view of this three-path workflow. It complements the system-level architecture in Figure~\ref{fig:system_architecture} by showing the construction-zone image, the vehicle and connectivity context exposed through MCP, and the driver-advisory, cooperative-message, and automation-intent outputs used in the prototype. The model logos illustrate a replaceable inference backend rather than a comparison between the displayed models; the reported on-road run used Qwen3-VL. Likewise, the connectivity branch represents semantic content proposed by the model. Deterministic validation, message selection, addition of system-derived fields, and serialization occur downstream as described below.

\subsubsection{Message-Intent Processing}

The model-generated \texttt{message\_intent} contains semantic fields such as event type, scene description, relevant road users or objects, recommended response, confidence, and validity information. Vehicle identity, geographic position, and timestamp are obtained from system sources rather than generated by the model. The prototype processing sequence is
\begin{equation}
\text{multimodal context}
\rightarrow
\text{message intent}
\rightarrow
\text{deterministic validation and encoding}
\rightarrow
\text{MQTT publication}.
\end{equation}

This division is necessary because a model can propose context-dependent semantic content but cannot guarantee message-schema conformance. The adapter, rather than the model, determines whether a supported mapping exists and whether the resulting payload is admissible for publication.

\subsubsection{MetaAction Command Bridge}

A MetaAction is a structured, parameterized request for an existing function in the Level-2 stack. The reported prototype exercises a speed-related MetaAction; it does not evaluate a general library of steering and longitudinal maneuvers. A representative output has the following form:

\begin{verbatim}
{
  "meta_action": {
    "name": "set_speed",
    "desired_speed_mps": 5.23,
    "confidence": 0.92,
    "validity_ms": 500,
    "max_rate_mps2": 1.5
  }
}
\end{verbatim}

The command bridge first validates the action name, data types, required parameters, permitted ranges, and validity period. It then checks whether the action is available for the active vehicle configuration and compatible with the current automation state. An accepted request is mapped to the applicable openpilot planning or driver-interface function; malformed, stale, unsupported, or out-of-range requests have no effect. The openpilot stack retains responsibility for trajectory generation, control tracking, actuator commands, driver monitoring, and vehicle-specific safety enforcement.

\begin{figure}[!htbp]
    \centering
    \includegraphics[width=0.98\textwidth]{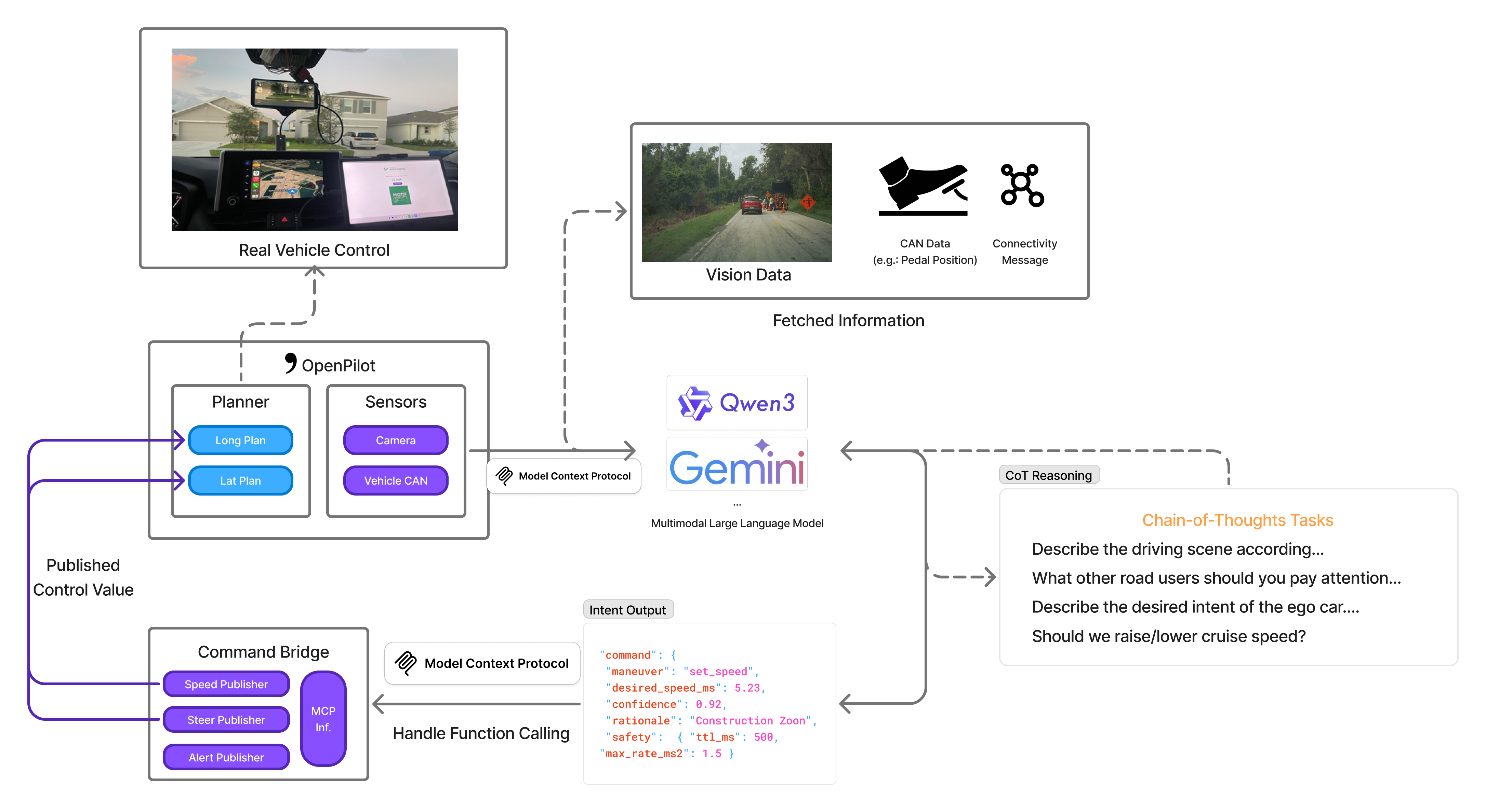}
    \caption{MetaAction command-bridge design for mapping validated, structured GenAI outputs to existing openpilot functions. The reported construction-zone experiment exercised only the speed-related path; the other publisher blocks illustrate extension points and are not evaluated in this study.}
    \label{fig:metaaction_bridge}
\end{figure}

Figure~\ref{fig:metaaction_bridge} details this separation. The model receives contextual information and proposes a structured high-level action, while the command bridge determines whether an admissible parameter may be passed to an existing openpilot function. The arrows therefore represent the path of an accepted high-level request, not direct model access to steering, braking, or the CAN bus. Although the diagram shows candidate speed, steering, and alert publisher interfaces in the extensible bridge design, the experiment reported here enabled and exercised only the speed-setting pathway.

\section{Discussion}

The results establish the functional feasibility of integrating an affordable retrofit OBU, lightweight connectivity, and GenAI with an existing open-source Level-2 automation stack. The contribution is therefore a minimum-but-functional platform for experimentation on supported production vehicles, rather than a demonstration of fleet-scale effectiveness or safety-critical CDA performance. This distinction is important when considering both the potential benefits of broader deployment and the functional boundaries introduced by reducing hardware and integration cost.

\subsection{Implications of Broader Scale-CDA Deployment}

The primary advantage of \tech is that it lowers the incremental cost and vehicle-integration effort required to add cooperative research functions to a production vehicle. A representative core OBU costs approximately \$820, and the OpenDBC/openpilot foundation provides reusable vehicle-state and supported Level-2 interfaces across more than 300 production-vehicle configurations. This inherited coverage does not mean that \tech has been validated on every configuration, but it provides a substantially larger starting point than a platform engineered for one experimental vehicle.

If Scale-CDA-equipped passenger vehicles become more numerous, the potential value of the platform could grow through a network effect. Each equipped vehicle can act as both an observation source and an information recipient. At greater penetration, the resulting fleet could provide wider spatial and temporal coverage of traffic conditions, road hazards, work zones, and vehicle-operation data. The same connectivity layer could distribute location-specific advisories or experimental traffic-management information to more road users without requiring each research vehicle to be built around a dedicated V2X radio and vehicle-specific control interface. These are prospective benefits: the present study does not measure market penetration, network effects, traffic-flow improvement, or safety outcomes.

Data collection is a useful secondary capability even when no cooperative action is requested. Supported CAN signals, forward-camera observations, positioning information, and received messages can be recorded through a common interface for transportation and vehicle research. Our OpenLKA work \citep{wang2025openlkaopendatasetlane} illustrates how the broader OpenDBC/openpilot toolchain can support the collection of production-vehicle driving data. The available measurements nevertheless depend on the vehicle, connection method, and supported signal definitions; Scale-CDA does not provide unrestricted access to every signal on every vehicle.

\subsection{Work-Zone Advisories, Variable Speed Limits, and Agency Applications}

Lightweight connectivity may be particularly useful for non-safety-critical agency applications such as work-zone information, advisory speeds, traffic-management notices, and experimental telemetry. An agency or roadside service could publish a location-specific advisory through the MQTT interface. After source, format, location, and freshness checks, an equipped vehicle could present the information to the driver or offer it to a supported, supervised Level-2 speed-setting function through the MetaAction bridge. The message would not directly command the vehicle or bypass the existing automation state, driver supervision, and vehicle-specific safety checks.

A concrete opportunity is integration with the U.S. DOT Work Zone Data Exchange (WZDx) program. WZDx enables infrastructure owners and operators to publish harmonized work-zone data for use by third parties, with the goal of making information about work-zone activity available to vehicles, automated-driving systems, and human drivers \citep{fhwa_wzdx}. Scale-CDA could serve as a vehicle-facing delivery layer for these feeds. A cloud or roadside service could ingest a WZDx record, identify equipped vehicles approaching the affected road segment, translate the relevant location, lane-closure, worker-presence, and validity information into a supported advisory payload, and publish it through the Scale-CDA MQTT interface. The OBU could then display a work-zone warning or pass an admissible speed recommendation to the supervised MetaAction pathway. In this arrangement, WZDx supplies harmonized work-zone information, while Scale-CDA provides the last-mile connection to supported production vehicles; WZDx itself is not treated as a safety-critical vehicle-control channel.

The same lightweight pathway can be extended to variable speed limit (VSL) applications. A traffic-management center could publish a speed advisory together with its road segment, direction, source, issue time, and validity interval. After location and freshness checks, Scale-CDA could display the current advisory or, where the vehicle and automation state permit, propose the value through the validated \texttt{set\_speed} MetaAction. As the number of equipped passenger vehicles grows, WZDx-based warnings and VSL advisories could reach a broader portion of the traffic stream and provide agencies with a practical test bed for studying information coverage, driver response, and traffic-management effects. From a traffic-management perspective, broader and more timely delivery could help vehicles respond earlier to work zones, changing capacity, or emerging queues. These potential benefits have not yet been quantified by the present prototype.

\begin{figure}[!htbp]
    \centering
    \includegraphics[width=0.72\linewidth]{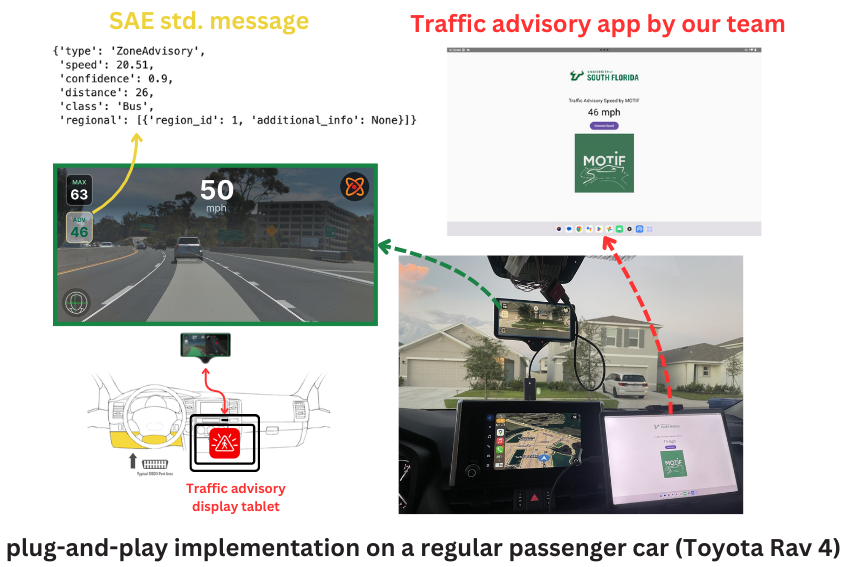}
    \caption{Illustrative advisory-speed application on a Toyota RAV4. The prototype demonstrates information delivery and display; safety-critical control and independent SAE-message conformance are outside the scope of this example.}
    \label{fig:advisory_demo}
\end{figure}

Figure~\ref{fig:advisory_demo} shows an example advisory-speed interface implemented for mobile and tablet displays. Such an interface offers a relatively accessible path for work-zone and VSL research pilots because it can provide information without requiring automated control. Integration with operational WZDx feeds, commercial infotainment systems, systematic usability testing, and evaluation of driver and traffic response remain future work. More demanding applications such as cooperative maneuver control, platooning, emergency intervention, or replacement of existing tolling systems would require capabilities and evidence beyond the current prototype.

\subsection{Cost--Function Tradeoffs and Prototype Boundaries}

The lower cost of Scale-CDA is achieved by reusing existing vehicle interfaces and Level-2 functions and by selecting commodity sensing, computing, and communication components. This choice expands accessibility but also defines the platform's current functional boundary. Table~\ref{tab:tradeoffs} summarizes the principal advantages and corresponding limitations.

\begin{table}[!htbp]
\centering
\caption{Principal cost--function tradeoffs in the Scale-CDA prototype.}
\label{tab:tradeoffs}
\begin{tabular}{@{}p{0.18\linewidth}p{0.44\linewidth}p{0.37\linewidth}@{}}
\toprule
\textbf{Dimension} & \textbf{Advantage of the minimum platform} & \textbf{Current boundary or limitation} \\
\midrule
Hardware & Commodity, replaceable components and an approximately \$820 core OBU & No redundant sensing, automotive-grade hardening, or safety certification of the assembled OBU \\
Vehicle interface & Reuses OpenDBC/openpilot definitions instead of rebuilding each OEM interface & Available signals and Level-2 functions vary by model and configuration; broad inherited coverage has not been validated by this study \\
Connectivity & MQTT over Wi-Fi or LTE reduces specialized-radio requirements for experimental applications & Only one low-speed Wi-Fi setting was measured; latency tails, loss, handoff, congestion, security, and LTE performance remain insufficiently characterized \\
GenAI & Combines camera, CAN, and received-message context and produces advisories, message intents, and high-level MetaActions & The road demonstration covers one representative scene; semantic correctness, robustness, latency, and safety benefit were not evaluated systematically \\
Local inference & Commodity edge devices can execute smaller VLMs and retain data onboard & The screened local configurations were not sufficiently responsive for the reported on-road pipeline, which used a cloud model \\
Message support & Deterministic adapters separate semantic generation from supported SAE J2735-based encoding & The full J2735 message dictionary and independent conformance testing are not implemented \\
\bottomrule
\end{tabular}
\end{table}

The lightweight connectivity result requires particular caution. The 5.25-ms mean RTT observed in the parking-lot experiment is encouraging for the evaluated telemetry and advisory messages, but approximately 2\% of observations exceeded 50~ms. The experiment does not establish bounded latency, reliability, coverage, or performance under congested and high-speed conditions. LTE was implemented as an alternative transport but was not benchmarked in the reported test. Consequently, the present evidence is insufficient to determine whether Scale-CDA could support safety-critical applications of the type targeted by dedicated V2X technologies. The evaluated prototype should be treated as a platform for supervised, non-safety-critical applications rather than as a replacement for DSRC or C-V2X sidelink communication.

Feasible GenAI integration also does not imply that model outputs are correct or safe. Schema, range, capability, and freshness checks can prevent malformed outputs from reaching a message or automation interface, but they cannot by themselves prove that a scene interpretation or recommended response is semantically correct. This is why GenAI remains outside the low-level control loop and the driver continues to supervise the Level-2 feature. Broader deployment would additionally raise cybersecurity, privacy, data-governance, and message-trust questions because more vehicles would produce and consume shared information and cloud inference may transmit selected context outside the vehicle.

\subsection{Research Needed to Move Beyond the Minimum Prototype}

Several evaluations are needed before the potential fleet-level implications can be established. Connectivity experiments should cover multiple environments, road speeds, traffic densities, payload sizes, and interference conditions while measuring packet loss, jitter, handoff behavior, MQTT quality-of-service settings, LTE performance, and end-to-end application deadlines. An agency pilot should additionally evaluate ingestion of operational WZDx feeds, road-segment matching, stale or conflicting records, VSL update timing, driver response, and resulting traffic effects. Vehicle testing should quantify installation effort, available signals, and supported automation functions across representative OEM configurations rather than infer uniform capability from ecosystem coverage.

To address the installation and cross-OEM deployment gap, we propose a pilot deployment of 100 Scale-CDA kits in vehicles owned by Tampa-area residents, subject to the required institutional and vehicle-safety approvals. In coordination with USDOT and local partners, the project would subsidize installation services at participating automotive service centers rather than require residents to install the equipment themselves. The research team would develop vehicle-specific procedures and train participating technicians to access the ADAS module, disconnect the OEM harness, insert the Scale-CDA harness, reassemble the interface, and complete diagnostic and functional checks. The pilot would quantify installation time, compatibility failures, technician errors, repeat-service requirements, system reliability, and participant acceptance across vehicle configurations. These measurements would establish whether a trained-service model can convert the current research installation into a safe, repeatable, and scalable pathway for deploying Scale-CDA within the existing vehicle fleet.

The GenAI pathway requires an annotated event set, vision-only and deterministic baselines, end-to-end latency measurements, semantic-correctness metrics, adversarial and stale-input tests, and failure injection at the message and MetaAction boundaries. Standards-oriented work should identify the supported SAE J2735 message types and conduct independent conformance testing. Finally, security controls, privacy policy, driver-interface usability, and the effects of increasing equipped-vehicle penetration should be evaluated before agency or fleet deployment. These steps would determine where the low-cost platform remains sufficient and where safety-critical radios, redundant sensors, higher-performance computing, or additional assurance mechanisms are necessary.

\section{Conclusion}

Scale-CDA demonstrates a feasible path for adding lightweight connectivity and GenAI research capabilities to supported production vehicles through an affordable retrofit OBU and an existing open-source Level-2 automation stack. A representative core configuration costs approximately \$820. In the moving-vehicle Wi-Fi experiment, the system achieved a mean message RTT of 5.25~ms and mean negotiated physical-layer rates of 98.51~Mb/s for transmission and 109.17~Mb/s for reception, although approximately 2\% of observations exceeded 50~ms. The GenAI prototype combined a camera observation, decoded CAN states, and received connectivity context and exercised driver-advisory, cooperative-message, and speed-related MetaAction pathways through deterministic adapters rather than direct packet or actuator access.

If deployed across more passenger vehicles, this low-cost approach could increase the number and diversity of vehicles available for cooperative data collection, WZDx-based work-zone warnings, VSL advisory delivery, and CDA experimentation. Its principal value is therefore not a new wireless standard, autonomous-driving controller, or foundation model, but a reusable integration layer that lowers the threshold for studying cooperative functions on heterogeneous production vehicles.

The reduction in cost comes with clear functional boundaries. The reported results are limited to one Wi-Fi setting, supported SAE J2735-based message mappings, exploratory local-inference benchmarks, and one representative GenAI road scene. Cross-OEM capability is inherited from OpenDBC/openpilot but was not validated across all supported configurations. Most importantly, this study does not determine whether Scale-CDA can meet the latency, reliability, security, and assurance requirements of safety-critical V2X applications. The current prototype should be used for supervised, non-safety-critical research while future work evaluates the conditions under which additional communication, sensing, computing, and safety mechanisms are required.


\bibliographystyle{plainnat}

\bibliography{ref}


\end{document}